\documentclass[a4paper,aps,showpacs]{revtex4}
\usepackage[reqno]{amsmath}
\usepackage{epsf,amsfonts,amssymb,dcolumn}
\newcommand{\ebox}[2]{\epsfxsize=#1 \epsfbox[10 30 560 590]{#2}}

\newcommand{\Dslash}{/\hspace{-1.5ex}D}
\newcommand{\Dslashbak}{\overleftarrow{\Dslash}}

\newcommand{\pslash}{\not\!p}

\newcommand{\kslash}{\not\!k}
\newcommand{\half}{\frac{1}{2}}
\newcommand{\quarter}{\frac{1}{4}}
\newcommand{\bra}{\langle}
\newcommand{\ket}{\rangle}
\newcommand{\Tr}{\operatorname{tr}}

\newcommand{\order}{\mathcal{O}}
\newcommand{\sss}{\scriptscriptstyle}

\newcommand{\khat}{\hat{k}}
\newcommand{\csw}{c_{\text{sw}}}
\newcommand{\psibar}{\overline{\psi}}
\newcommand{\mtil}{\widetilde{m}}

\newcommand{\zz}{Z^{(0)}}
\newcommand{\zmz}{Z^{(0)}_m}
\newcommand{\zmp}{Z^{\sss(+)}_m}
\newcommand{\zmpI}{Z^{\sss(+)}_{m,I}}
\newcommand{\zmpR}{Z^{\sss(+)}_{m,R}}
\newcommand{\dmz}{\Delta M^{(0)}}
\newcommand{\dmp}{\Delta M^{\sss(+)}}
\newcommand{\dmm}{\Delta M^{\sss(-)}}
\newcommand{\err}[2]{\raisebox{-0.4ex}
{$\stackrel{\scriptstyle +#1}{\scriptstyle -#2}$}}

\newcommand{\prev}{\cite{Skullerud:2000un}}
\newcommand{\UKQCD}{\cite{Bowler:1999ae}}
\begin{document}

\preprint{ADP-01-07/T443 \hskip0.5cm
          DESY 01-022}

\title{Nonperturbative Improvement and Tree-level Correction of the \\
Quark Propagator} 

\author{Jonivar Skullerud}
\email{jonivar@mail.desy.de}
\homepage{http://www.bigfoot.com/~jonivar/}
\affiliation{DESY Theory Group, Notkestra{\ss}e 85, D--22603 Hamburg, Germany}
\author{Derek B.\ Leinweber}
\email{dleinweb@physics.adelaide.edu.au}
\homepage{http://www.physics.adelaide.edu.au/~dleinweb/}
\author{Anthony G.\ Williams}
\email{awilliam@physics.adelaide.edu.au}
\homepage{http://www.physics.adelaide.edu.au/cssm/}
\affiliation{CSSM and Department of Physics and Mathematical Physics, \\
Adelaide University, Australia 5005}

\begin{abstract}
We extend an earlier study of the Landau gauge quark propagator
in quenched QCD where we used two forms of the $\order(a)$-improved
propagator with the
Sheikholeslami-Wohlert quark action.
In the present study we use the nonperturbative value for the clover
coefficient $\csw$ and mean-field 
improvement coefficients in our improved quark propagators.  We
compare this to our earlier results which used the mean-field $\csw$
and tree-level improvement
coefficients for the propagator.  We also compare three different
implementations of tree-level correction:
additive, multiplicative, and hybrid.  We show that the hybrid approach
is the most robust and reliable and can successfully deal even with
strong ultraviolet behavior and zero-crossing of the lattice
tree-level expression.  We find good agreement between our improved
quark propagators when using the appropriate nonperturbative improvement
coefficients and hybrid tree-level correction.  We also present
a simple extrapolation of the quark mass function to the chiral
limit.
\end{abstract}


\maketitle

\section{Introduction}
\label{Sec:intro}

Lattice studies of the quark propagator provide a direct,
model-independent window into the
mechanism of dynamical chiral symmetry breaking and its momentum
dependence.  In addition it provides insight into the the nature and
location of the transition region of QCD where inherently nonperturbative
behavior evolves into the more analytically accessible perturbative form.
Furthermore, direct lattice calculations
of the quark propagator inform hadron model building from the relativized
constituent quark picture to quark models based on Schwinger-Dyson
equations \cite{Roberts:1994dr,Alkofer:2000wg}.

In a recent paper \cite{Skullerud:2000un} we presented a method for
removing the dominant ultraviolet tree-level lattice artifacts in the
momentum-space quark propagator.  This was a generalization of
the concept of tree-level correction, which was first introduced
in the study of the gluon propagator
\cite{Leinweber:1998im,Leinweber:1998uu,Bonnet:2000kw,Bonnet:2001uh}. 
It was shown that, for two different
$\order(a)$-improved propagators $S_I$ and $S_R$, see
Eqs.~(\ref{eq:S-imp}) and (\ref{eq:S-rot}), and using a mean-field improved
action, this leads to a dramatic improvement in the ultraviolet
behavior of the propagator.  However, the remaining ultraviolet
artifacts are sufficiently large to make the results unreliable beyond
$pa\sim1.2$.  Moreover, the two improved propagators remain
discernibly different
even in the infrared, yielding different estimates for the infrared
quark mass.

Here we will present results using nonperturbatively determined
values for the $\order(a)$ improvement coefficients, rather than the
tree-level and mean-field improved coefficients used in
Ref.~\cite{Skullerud:2000un}.  We will also present two alternative
techniques for removing tree-level artifacts and will discuss the
relative merits of the three methods.

\section{Improvement}
\label{Sec:improve}

The general scheme for $\order(a)$ improvement of the quark propagator
was discussed in Ref.~\cite{Skullerud:2000un}.  Here we restrict ourselves
to presenting the formulae and definitions which we will be using in
this paper.  For further details, see Ref.~\cite{Skullerud:2000un} and
references therein.

The SW fermion action,
\begin{equation}
{\mathcal L}(x) = {\mathcal L}^W(x) 
- \frac{i}{4}\csw a\psibar(x)\sigma_{\mu\nu}F_{\mu\nu}(x)\psi(x) \, ,
\label{eq:SW}
\end{equation}
combined with appropriate improvements of operators can be shown
\cite{Luscher:1996sc} to remove all $\order(a)$ errors in on-shell
matrix elements.  For off-shell quantities such as the quark
propagator it is not that simple, and no general proof of $\order(a)$
improvement is known.  Indeed, to calculate gauge dependent quantities
one might expect to have to introduce gauge non-invariant (but BRST
invariant) terms in the action.  However, at tree level it is possible
to proceed by adding all possible dimension-5 operators to the action
and eliminating all but the clover (SW) term by a field redefinition
\cite{Heatlie:1991kg}.  Beyond tree level, one may proceed by adding
all possible terms with the correct dimensionality and quantum numbers
to the operator in question, and tuning the parameters to eliminate
$\order(a)$ terms.  Ignoring the gauge non-invariant terms (which are
discussed in Refs.~\cite{Dawson:1997gp,Becirevic:1999kb}) we may write down
the following expressions for the $\order(a)$ improved quark
propagator,
\begin{align}
S_I(x,y) & \equiv \bra S_I(x,y;U)\ket 
\equiv \bra(1+b_q am)S_0(x,y;U) - a\lambda\delta(x-y)\ket \, ,
\label{eq:S-imp} \\
S_R(x,y) & \equiv \bra S_R(x,y;U)\ket
\equiv \bra (1+b'_q am)\left[1-c'_q\Dslash(x)\right]
S_0(x,y;U) \left[1+c'_q\Dslashbak(y)\right] \ket \; .
\label{eq:S-rot}
\end{align}
where $S_0(x,y;U)$ for a given configuration $U$ is the inverse of the
fermion matrix.  Note that if we are only interested in on-shell
quantities such as hadronic matrix elements, the $\delta$-function can
be ignored, so we only need $S_0$ together with the improvement
coefficients for the various operators.  From this, $S_I$ is easily
obtained, whereas $S_R$ is computationally somewhat more expensive.

The coefficients $b_q$, $b'_q$, $\lambda$ and $c'_q$ must be tuned in
order to eliminate $\order(a)$ errors in the propagator (as far as
this is possible).  At tree level, their values are $b_q=1,
\lambda=b'_q=\half, c'_q=\frac{1}{4}$.  The values for $b_q$ and
$\lambda$ have recently been calculated at one-loop level
\cite{Capitani:2000xi}.  The mean-field improved values for all these
coefficients may be obtained by dividing the tree-level values by the
mean link $u_0$.  The one-loop mean-field improved values have also
been calculated in Ref.~\cite{Capitani:2000xi} but as we will argue,
the small changes in values obtained by including the one-loop
contribution have very little practical effect, so we will not use
these here.  It should also be noted that the mean-field improved
value for $\lambda$ is very close to the nonperturbative value
reported in Ref.~\cite{Capitani:1997nr}.  This indicates that
mean-field improvement of the coefficients $b_q$ and $\lambda$ (or,
alternatively, $b'_q$ and $c'_q$) may be sufficient to remove
$\order(a)$ errors to the desired precision.

The bare mass also receives an $\order(a)$ correction, which can be
expressed as follows
\begin{equation}
\mtil = (1 + b_m am)m \qquad , \qquad 
am = \frac{1}{2\kappa}-\frac{1}{2\kappa_c} \; .
\label{eq:mimp}
\end{equation}
The coefficient $b_m$ has been calculated at one-loop order
\cite{Sint:1997jx},
\begin{equation}
b_m = -\half - 0.0962g_0^2 + \order(g_0^4) \; .
\label{eq:bm}
\end{equation}
When evaluating Eq.~(\ref{eq:bm}), we will be using the boosted coupling
constant $g^2=g_0^2/u_0^4$.

\section{Tree-level correction}
\label{Sec:correct}

In the continuum, the spin and Lorentz structure of the quark
propagator, together with parity symmetry, determines that the
propagator must have the following form,
\begin{equation}
S(\mu;p) = \frac{Z(\mu;p^2)}{i\pslash + M(p^2)}\equiv 
     \frac{1}{i\pslash A(\mu;p^2) + B(\mu;p^2)} \;.
\label{eq:generic-prop}
\end{equation}

On the lattice, what we measure is the bare (regularized but
unrenormalized)
propagator.  This differs from the renormalized propagator
in Eq.~(\ref{eq:generic-prop}) by an overall renormalization constant
$Z_2(\mu,a)$, which we will absorb into $Z(p)$, as we did in Ref.~\prev\, to
simplify the presentation of our results.

In Ref.~\prev\ we defined a tree-level correction
procedure involving an overall multiplicative correction and an
additive correction of the mass function, as follows,
\begin{equation}
  S^{-1}(pa) = \frac{1}{Z(pa)\zz(pa)} 
  \left[ia\kslash + aM^a(pa) + a\dmz(pa) \right] ,
\label{eq:additive}
\end{equation}
where $k_\mu=\sin(p_\mu a)/a$ and $\zz$ and $\dmz$ are defined by the
tree-level quark propagator,
\begin{equation}
\left(S^{(0)}(pa)\right)^{-1} = \frac{1}{\zz(pa)}
 \left[ i\kslash a + am + a\dmz(pa)\right] \, .
\end{equation}
The functions $Z(pa)$ and $M^a(pa)$ should then express the
nonperturbative behavior of the quark propagator, with the dominant
lattice artifacts removed.  We saw that this procedure led to a
dramatic improvement in the behavior of these functions, but at large
momenta the data could still not be trusted because of large
cancellations. 

Here we will consider an alternative, purely multiplicative tree-level
correction procedure, defined by
\begin{equation}
  S^{-1}(pa) = \frac{1}{Z(pa)\zz(pa)} 
  \left[ia\kslash + aM^m(pa)\zmz(pa) \right] ,
\label{eq:multiplicative}
\end{equation}
where $\zmz(pa)$ is defined by the tree-level expression,
\begin{equation}
\left(S^{(0)}(pa)\right)^{-1} = \frac{1}{\zz(pa)}
 \left[ i\kslash a + am\zmz(pa)\right] \, .
\end{equation}
The tree-level corrected mass function $M$ is thus obtained from the
uncorrected function $M^L\equiv\Tr S^{-1}/4N_c$ via
\begin{equation}
aM^m(pa) = M^L(pa)/\zmz(pa) \, .
\label{eq:multip-correct}
\end{equation}
This procedure should not suffer from the problem of large
cancellations.  However, it will encounter problems when either the
numerator or denominator of Eq.~(\ref{eq:multip-correct}) crosses or is
close to zero.  In order to remedy this problem, we consider a
third, `hybrid'
scheme, where the negative part of the tree-level expression is
subtracted, while the remaining positive part is multiplicatively
corrected.  Specifically, we define $\dmp, \dmm$ such that
\begin{gather}
\dmp(pa) + \dmm(pa) = \dmz(pa)
\label{eq:hybrid-pmdef1} \\
\dmp(pa) \geq 0 ;\quad \dmm(pa) \leq 0 \quad \forall pa
\label{eq:hybrid-pmdef2}
\end{gather}
Then we can write
\begin{equation}
M^{(0)}(pa) = am + \dmp(pa) + \dmm(pa) \equiv am \zmp(pa) + \dmm(pa)
\, .
\label{eq:hybrid-def}
\end{equation}
The tree-level corrected mass function $M^h(pa)$ is then
\begin{equation}
aM^h(pa) = \left(M^L(pa)-a\dmm(pa)\right)/\zmp(pa) \, .
\end{equation}
The definition of this scheme contains an ambiguity, since it is
obvious that we may still satisfy (\ref{eq:hybrid-pmdef1}),
(\ref{eq:hybrid-pmdef2}) by adding any strictly positive term to
$\dmp$ and subtracting the same term from $\dmm$; e.g.\ by taking
$\dmp\to\dmp+k^2; \dmm\to\dmm-k^2$.  In order to remove this
ambiguity, we add the following criteria:
\begin{enumerate}
\item Factoring out a common (positive) denominator, $\dmp$ and $\dmm$
should be polynomials in the 4 variables $k^2, \khat^2, \Delta k^2$
and $m$,
where we have defined
\begin{equation}
\khat_\mu 
 \equiv  \frac{2}{a}\sin(p_{\mu}a/2) ; \qquad
a^2\Delta k^2 \equiv  \khat^2 - k^2  \, ;
\end{equation}
\item The coefficient of each term must be positive for $\dmp$ and
negative for $\dmm$;
\item Any one monomial in $k^2, \khat^2, \Delta k^2$ and $m$ can only
occur in one of $\dmp$ or $\dmm$; eg., if there is a term proportional
to $mk^2$ in $\dmm$ there cannot be a term proportional to $mk^2$ in
$\dmp$
\end{enumerate}
These criteria ensure that $\dmp$ and $\dmm$ are as small as
possible, leading to the minimum possible distortion of the data.

Specifically, the expressions we use are
\begin{align}
\zmpI(p) & = 
 \frac{am + (b_q-\lambda)a^2m^2 + \lambda a^4\Delta k^2 
   + (\half b_q-\lambda)a^3m\khat^2}{am(1+am)} \, ,
\label{eq:zmp-imp} \\
a\dmm_I(p) & 
  =  -\frac{\lambda a^4\khat^4/2 + (2\lambda-1)a^2\khat^2}{2(1+am)} 
\, , \label{eq:dmm-imp} \\
a\zmpR(p) & =
\frac{1}{amA'_R(p)}\Bigl(am + \half a^4\Delta k^2\Bigr) \, ,
\label{eq:zmp-rot} \\
a\dmm_R(p) &=
-\frac{1}{16A'_R(p)}\Bigl(a^3mk^2 + \half a^4k^2\khat^2\Bigr) \, ,
\label{eq:dmm-rot} \\
\intertext{where we have written}
A'_R(p) & = 1 + \half am + \frac{3}{16}a^2k^2 + \quarter a^4\Delta k^2
\, . \\
\end{align}

It should be remembered that the mass function $M(p)$ must be
renormalization-point independent in a renormalizable theory and that
the current quark mass at the renormalization point $m(\mu)$ is given
by $m(\mu)=M(p=\mu)$.  The ultraviolet mass function is of course only
constant up to logarithmic corrections.  The multiplicative and hybrid
tree-level correction ensures that the zeroth-order perturbative
behavior of the mass function in the ultraviolet matches that of the
continuum.  The logarithmic corrections should principle show up in
the lattice data, as they did for the gluon propagator in
Ref.~\cite{Leinweber:1998uu}.  However, this is a small effect
compared to the tree-level lattice artifacts.  It will be a measure of
the success of our improvement and correction scheme whether the
logarithmic corrections may be extracted from the lattice data.

\section{Results}
\label{Sec:results}

In addition to the data used in Ref.~\cite{Skullerud:2000un}, we have
analyzed data at $\beta=6.0$ using the nonperturbatively determined
value for $\csw$ (=1.769), and at $\beta=6.2$ using the mean-field
improved $\csw$ (=1.442).  This is a subset of the UKQCD data analyzed
in Ref.~\UKQCD, with the values for $\csw$ taken from
Ref.~\cite{Luscher:1997ug}.  The simulation parameters are given in
Table~\ref{tab:sim-params}.  Note that the values of $am$ are
different from those given in Ref.~\prev; this is because we have here
used the determination of $\kappa_c$ reported in Ref.~\UKQCD\ instead
of an earlier, preliminary value.  All the data shown have been
obtained from the raw data using the cylinder cut described in
Ref.~\prev.  The scale is taken from the hadronic radius $r_0$
\cite{Sommer:1994ce} using the interpolating formula of
Ref.~\cite{Guagnelli:1998ud} and the phenomenological value $r_0=0.5$
fm.  Note that this differs from the scale used in Ref.~\prev, which
was taken from an earlier determination of the string tension.  The
gauge fixing is identical to that of Ref.~\cite{Leinweber:1998uu},
which was also used in Ref.~\prev.  This is a version of lattice
Landau gauge that contains Gribov copies; the effects of these have
not been studied here.

\begin{table}
\begin{tabular}{cccccccrrr}
$\beta$ & Volume & $a^{-1}$ (GeV) & $\csw$ & $\kappa$ & prop & $am$ 
 & $m$ (MeV) & $\mtil$ (MeV) & $N_{\text{cfg}}$ \\ \hline
6.0 & $16^3\times48$ & 2.120 
    & MF & 0.13700 & $S_I$ & 0.0579 & 123 & 118 & 499 \\
& & &    &         & $S_R$ &        &     &     &  20 \\
& & &    & 0.13810 & $S_I$ & 0.0289 &  61 &  60 & 499 \\ \hline
6.0 & $16^3\times48$ & 2.120 
    & NP & 0.13344 & $S_I$ & 0.0498 & 105 & 102 &  10 \\
& & &    &         & $S_R$ &        &     &     &  20 \\
& & &    & 0.13417 & $S_I$ & 0.0294 &  62 &  61 &  10 \\
& & &    & 0.13455 & $S_I$ & 0.0188 &  40 &  39 &  10 \\ \hline
6.2 & $24^3\times48$ & 2.907
    & MF & 0.13640 & $S_I$ & 0.0399 & 116 & 113 &  54 \\
& & &    & 0.13710 & $S_I$ & 0.0212 &  62 &  61 &  54 \\ \hline
\end{tabular}
\caption{Simulation parameters.  `MF' and `NP' refer to the mean-field
improved and nonperturbatively determined values for $\csw$
respectively.  The improved propagators $S_I$ and $S_R$ are defined in
Eqs.\ (\ref{eq:S-imp}) and (\ref{eq:S-rot}).}
\label{tab:sim-params}
\end{table}

\subsection{Results with mean-field improved $\csw$}

\begin{figure}
\begin{center}
\setlength{\unitlength}{1.1cm}
\setlength{\fboxsep}{0cm}
\begin{picture}(14,7)
\put(0,0){\begin{picture}(7,7)\put(-0.9,-0.4){\ebox{8cm}{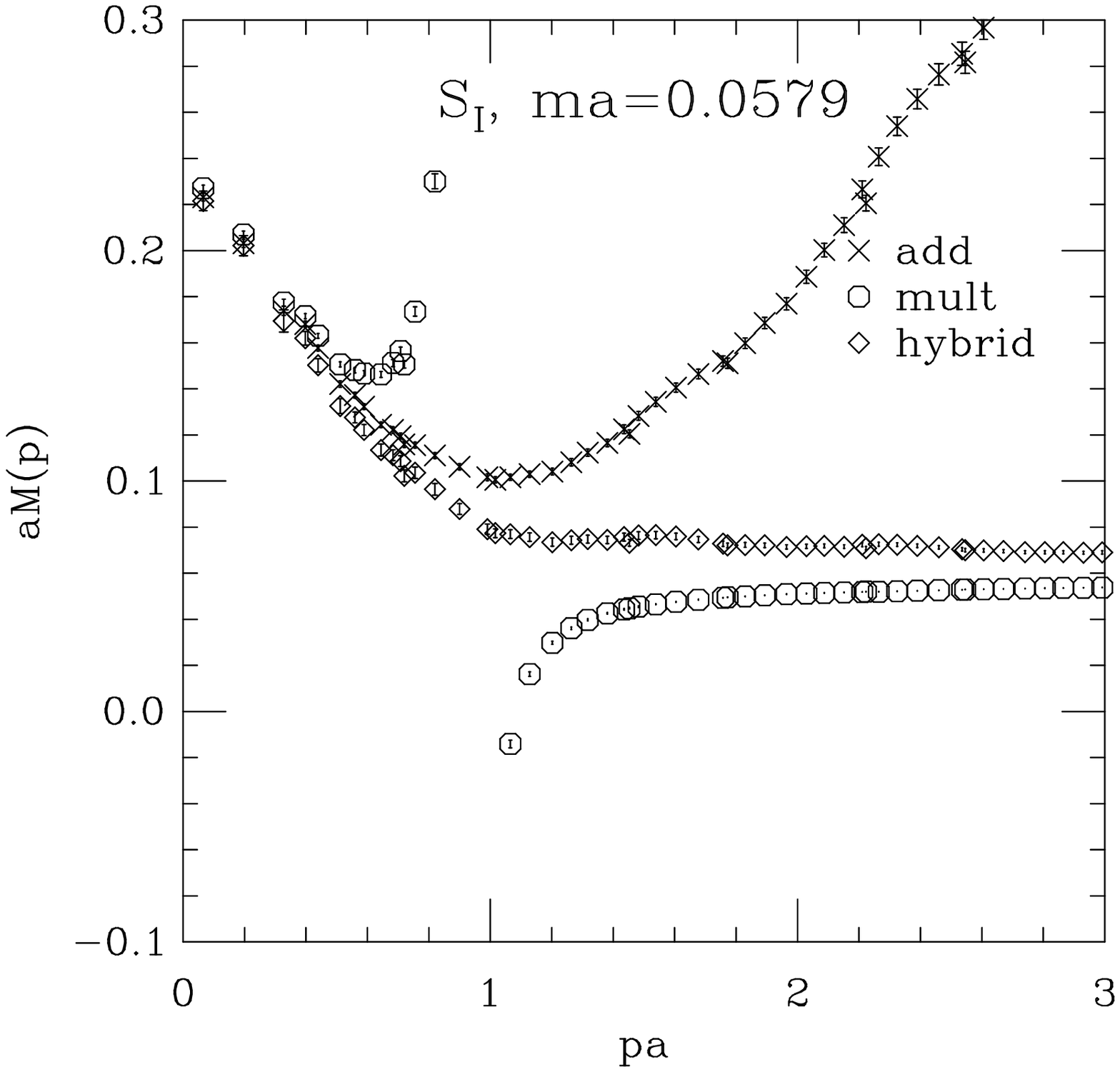}}\end{picture}}
\put(7,0){\begin{picture}(7,7)\put(-0.9,-0.4){\ebox{8cm}{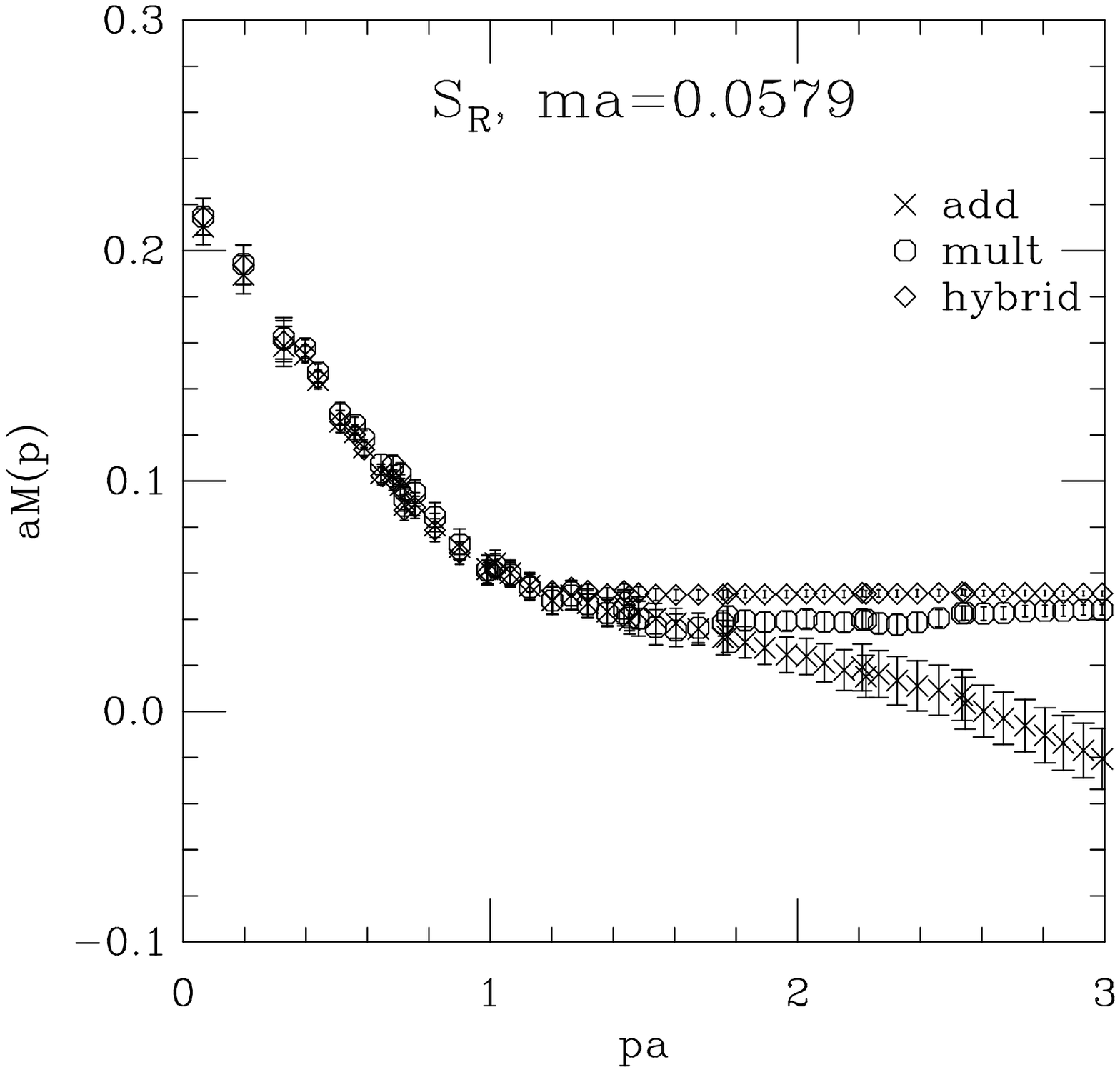}}\end{picture}}
\end{picture}
\end{center}
\caption{ The tree-level corrected mass function, for $S_I$ (left) and
$S_R$ (right) with $\csw=\text{MF}, \kappa=0.13700$, using additive,
multiplicative and hybrid correction.  The hybrid scheme is robust even in
the presence of a zero crossing of the tree-level mass function for
$S_I$.}
\label{fig:M-scheme-comp}
\end{figure}

We first consider the effect of employing the multiplicative and
hybrid correction schemes on the data analyzed in Ref.~\prev.
Figure~\ref{fig:M-scheme-comp} shows the tree-level corrected mass
function $M$ evaluated using the three schemes, for both $S_I$ and
$S_R$ at $\beta=6.0, \csw=\text{MF}$.  It is clear that the
ultraviolet behavior is much improved, but the multiplicatively
corrected $M$ from $S_I$ exhibits pathological behavior at
intermediate momenta.  This is a consequence of a zero crossing in the
tree-level mass function, leading to division by near-zero numbers in
Eq.~(\ref{eq:multip-correct}).  In the hybrid scheme, this
problem is absent.

It is worth noting that although the mass function approaches the
subtracted bare mass $m$ in the ultraviolet, the actual values
obtained using the multiplicative and hybrid schemes differ from each
other and from the bare mass by up to 20\%.  It is clear that at this
stage this procedure is not good enough to yield a good estimate of
current quark masses.  We also see that there is no sign in these data
of the logarithmic running of the current quark mass.

For $S_R$, we also see a clear
improvement in the ultraviolet behavior, as well as a small but
significant difference in the ultraviolet mass between the
multiplicative and hybrid schemes.  Since the tree-level mass function
for $S_R$ does not have any zero crossings, the multiplicatively
corrected mass does not exhibit the same pathological behavior as for
$S_I$.  The tree-level mass function for $S_R$ does not cross
zero, but does approach it, and the effects of this may be detected at
intermediate momenta.  Hence we consider the hybrid correction
to be more reliable for $S_R$. 

\subsection{Results with nonperturbative $\csw$}

\begin{figure}
\begin{center}
\setlength{\unitlength}{1.1cm}
\setlength{\fboxsep}{0cm}
\begin{picture}(14,7)
\put(0,0){\begin{picture}(7,7)\put(-0.9,-0.4){\ebox{8cm}{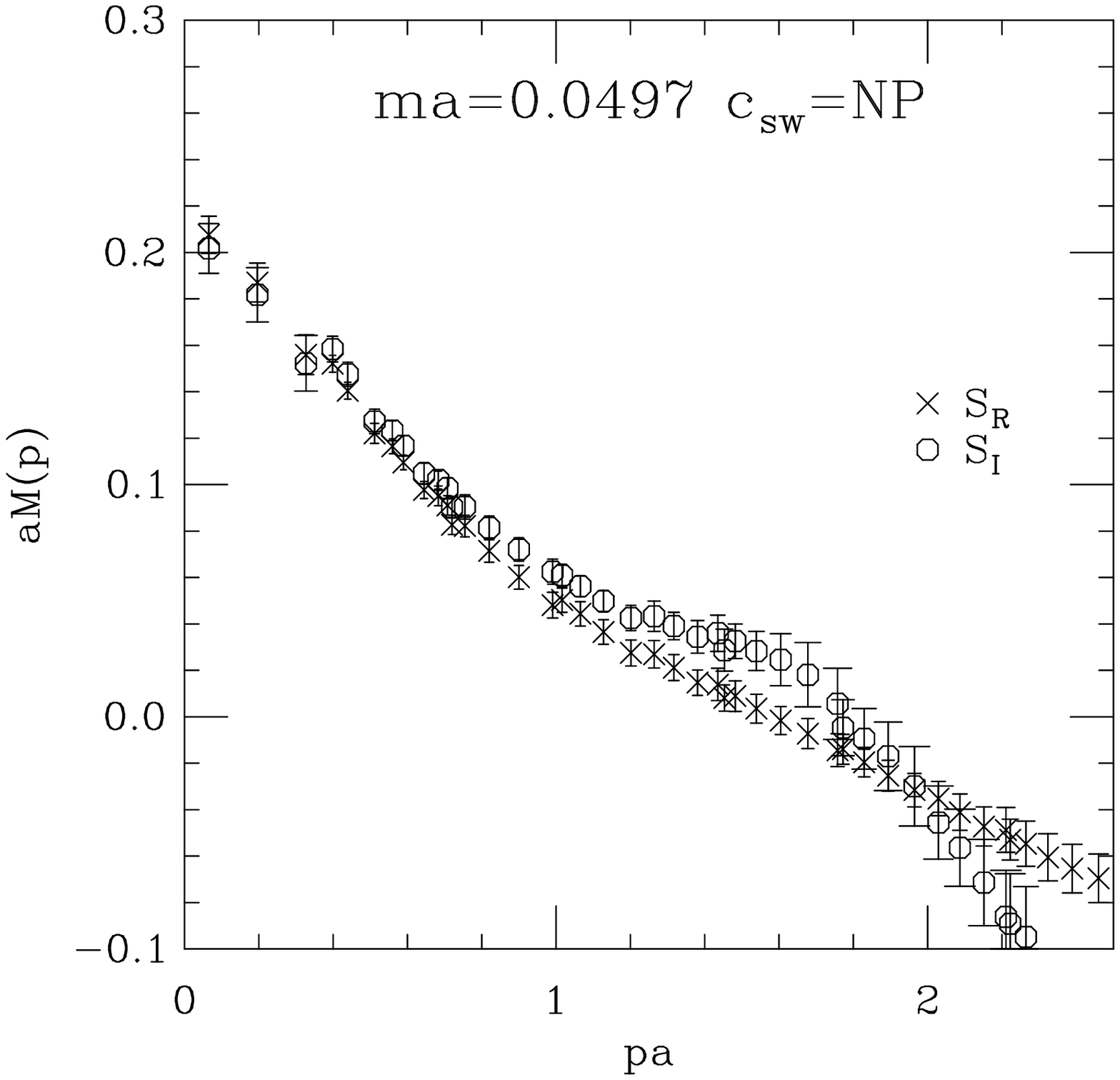}}\end{picture}}
\put(7,0){\begin{picture}(7,7)\put(-0.9,-0.4){\ebox{8cm}{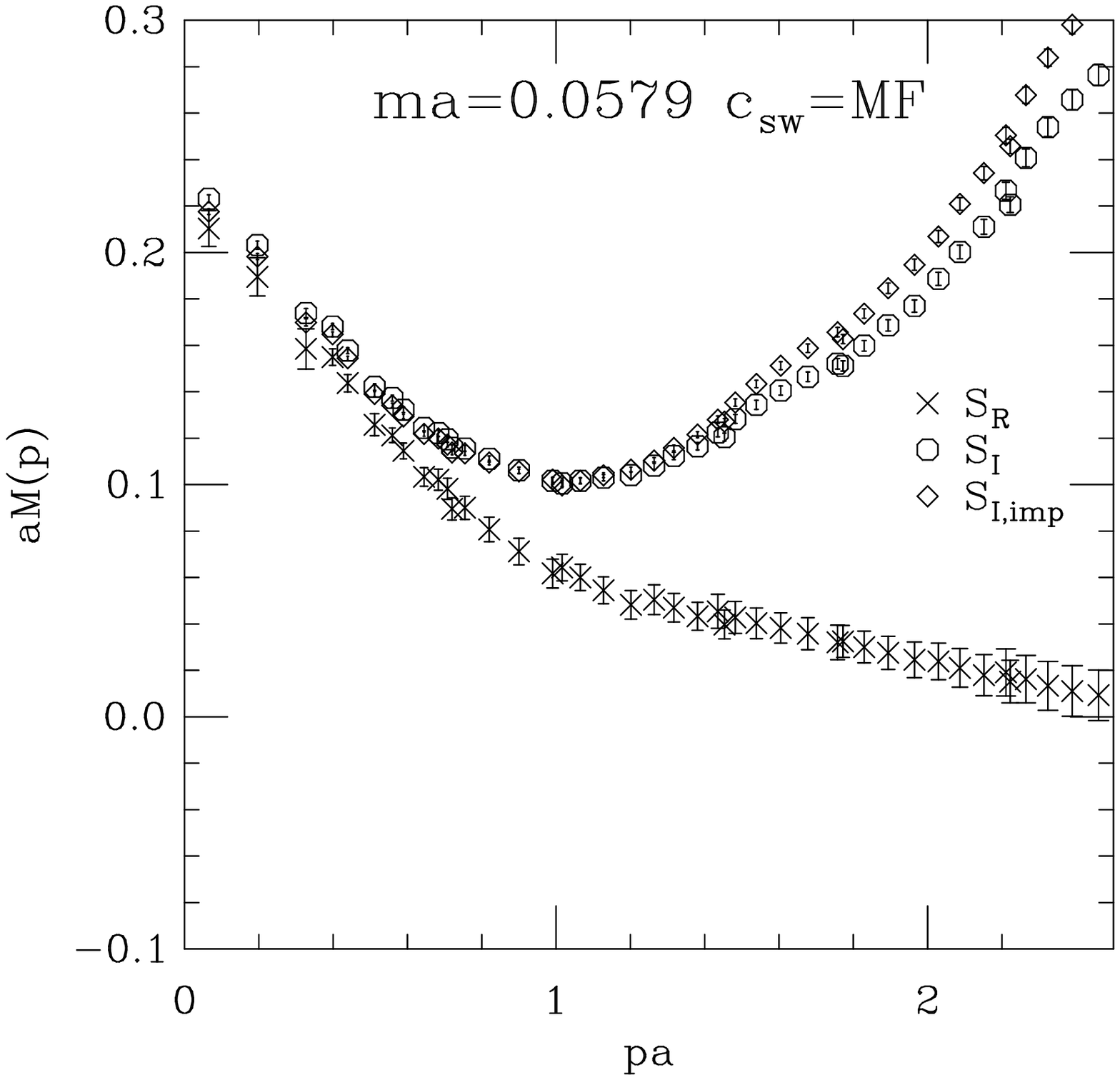}}\end{picture}}
\end{picture}
\end{center}
\caption{
The additively tree-level corrected mass function, for
$\csw=\text{NP}, \kappa=0.13344$
(left) and $\csw=\text{MF}, \kappa=0.13700$ (right).  $S_{I,\text{imp}}$ in
the right-hand figure is obtained using the mean-field improved rather
than the tree-level values for $b_q$ and $\lambda$ in
Eq.~(\protect\ref{eq:S-imp}). 
}
\label{fig:m_np_addcompare}
\end{figure}

We now turn to the effect of using nonperturbatively determined,
rather than tree-level or mean-field, improvement coefficients.
Figure \ref{fig:m_np_addcompare} shows, on the left, the mass function
obtained from our two improved propagators using the nonperturbative
value for $\csw$ and the mean-field values for $b_q, b'_q$ and $\lambda$ (we
have not been in a position to obtain data using mean-field or
nonperturbative values for $c'_q$, only the tree-level value
$c'_q=\quarter$).  As indicated in Sec.~\ref{Sec:improve}, the
nonperturbative values for the latter coefficients are currently not
known, but at least the nonperturbative value for $\lambda$ reported
in Ref.~\cite{Capitani:1997nr} is close to the mean-field value, and
it seems reasonable to guess that this is the case for the other
coefficients as well.  We therefore assume that although it is not
entirely consistent, it is not unreasonable to use the mean-field
values.  On the right of Fig.~\ref{fig:m_np_addcompare} are shown
equivalent data from Ref.~\prev, 
along with data using mean-field values for all the improvement
coefficients in $S_I$.

It is immediately clear that using the NP value for $\csw$ removes the
large discrepancy between the two improved propagators for
$pa\lesssim2$, even when using additive tree-level correction.  For
larger $pa$, the discrepancy remains, which is not unexpected since at
these momenta $\order(a^2)$ and higher errors become dominant.

\begin{figure}
\begin{center}
\setlength{\unitlength}{1.1cm}
\setlength{\fboxsep}{0cm}
\begin{picture}(14,7)
\put(3.5,0){\begin{picture}(7,7)\put(-0.9,-0.4){\ebox{8cm}{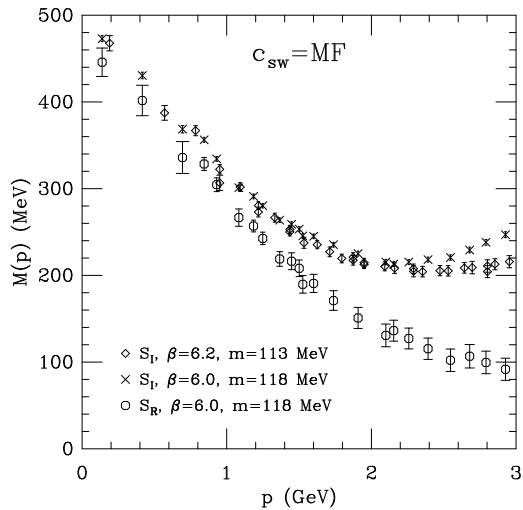}}\end{picture}}
\end{picture}
\end{center}
\caption{
The additively tree-level corrected mass function, for
$\csw=\text{MF}$, at $\beta=6.2$ and $\beta=6.0$.
}
\label{fig:m_beta_addcompare}
\end{figure}

It is instructive to compare
this with the effect of reducing the lattice spacing, as shown in
Fig.~\ref{fig:m_beta_addcompare}.  We do not have data for $S_R$ at
$\beta=6.2$, but we see that $S_I$ changes very little with $\beta$ in
the intermediate momentum range where the discrepancy becomes large.
Assuming that $S_R$ changes by a similar amount, we may conclude that
reducing the lattice spacing does very little to reduce the
discrepancy, although the behavior of $S_I$ at large momenta is
somewhat improved, as one would expect.  We can also see that going from
tree-level to mean-field values for the coefficients $b_q$ and $\lambda$
has only a very small effect.  We may also conclude that using
one-loop values for these coefficients will have negligible effect,
since the difference between tree-level and one-loop coefficients is
even smaller than that between the tree-level and mean-field improved
values.  This also gives us added confidence in the use of mean-field
rather than nonperturbative values for these coefficients.

\begin{figure}
\begin{center}
\setlength{\unitlength}{1.1cm}
\setlength{\fboxsep}{0cm}
\begin{picture}(14,14)
\put(0,7){\begin{picture}(7,7)\put(-0.9,-0.4){\ebox{8cm}{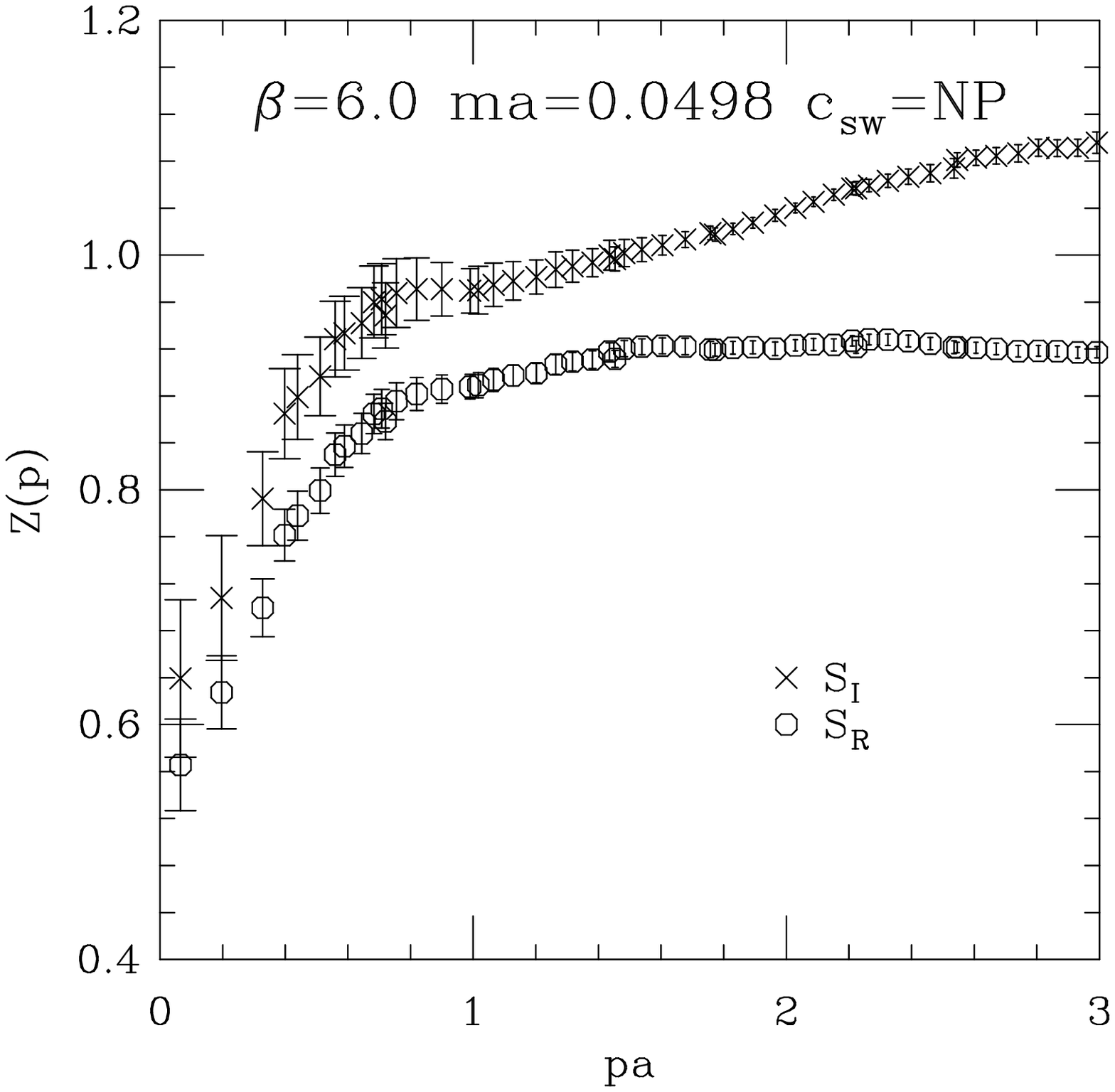}}\end{picture}}
\put(7,7){\begin{picture}(7,7)\put(-0.9,-0.4){\ebox{8cm}{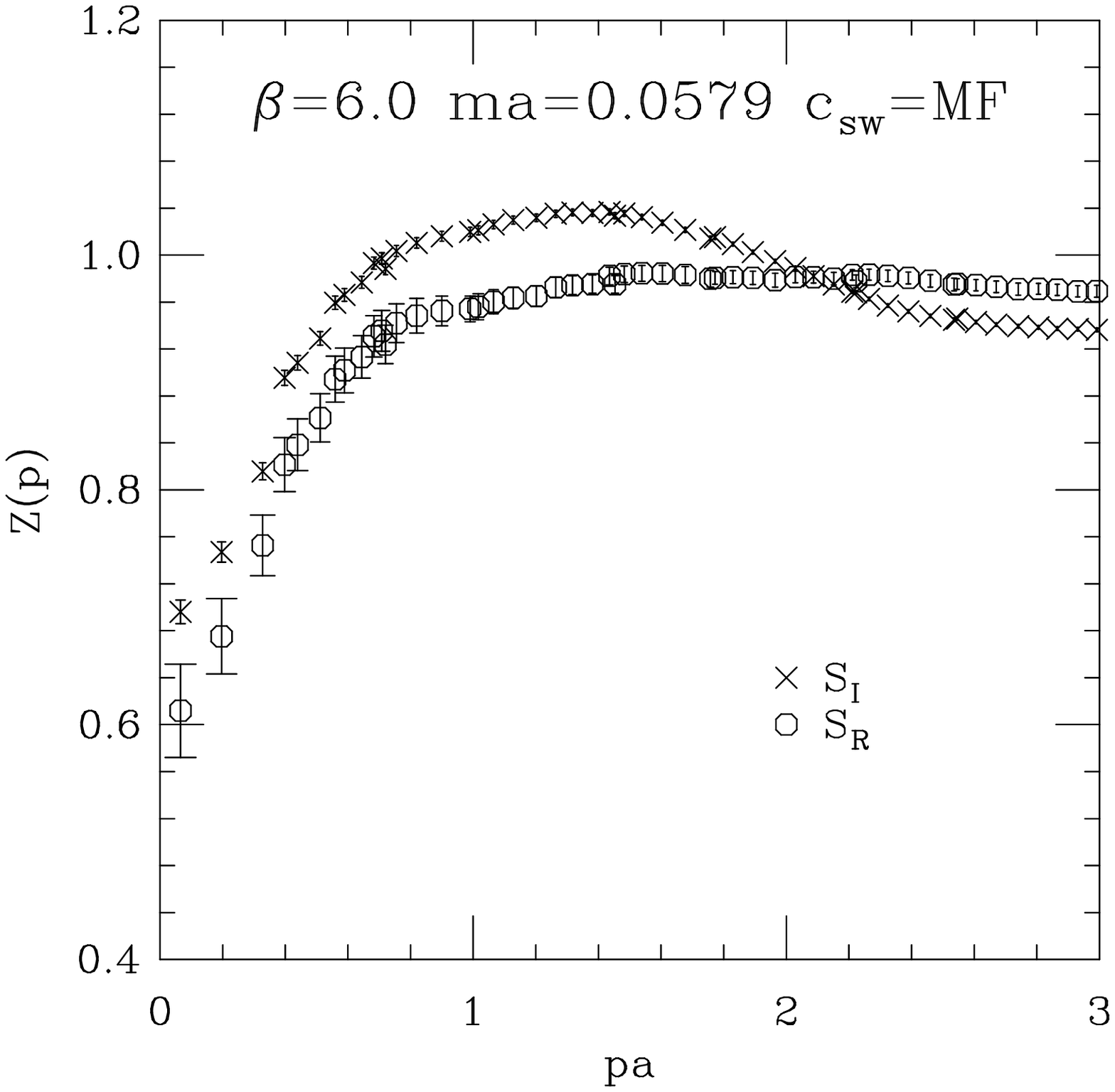}}\end{picture}}
\put(0,0){\begin{picture}(7,7)\put(-0.9,-0.4){\ebox{8cm}{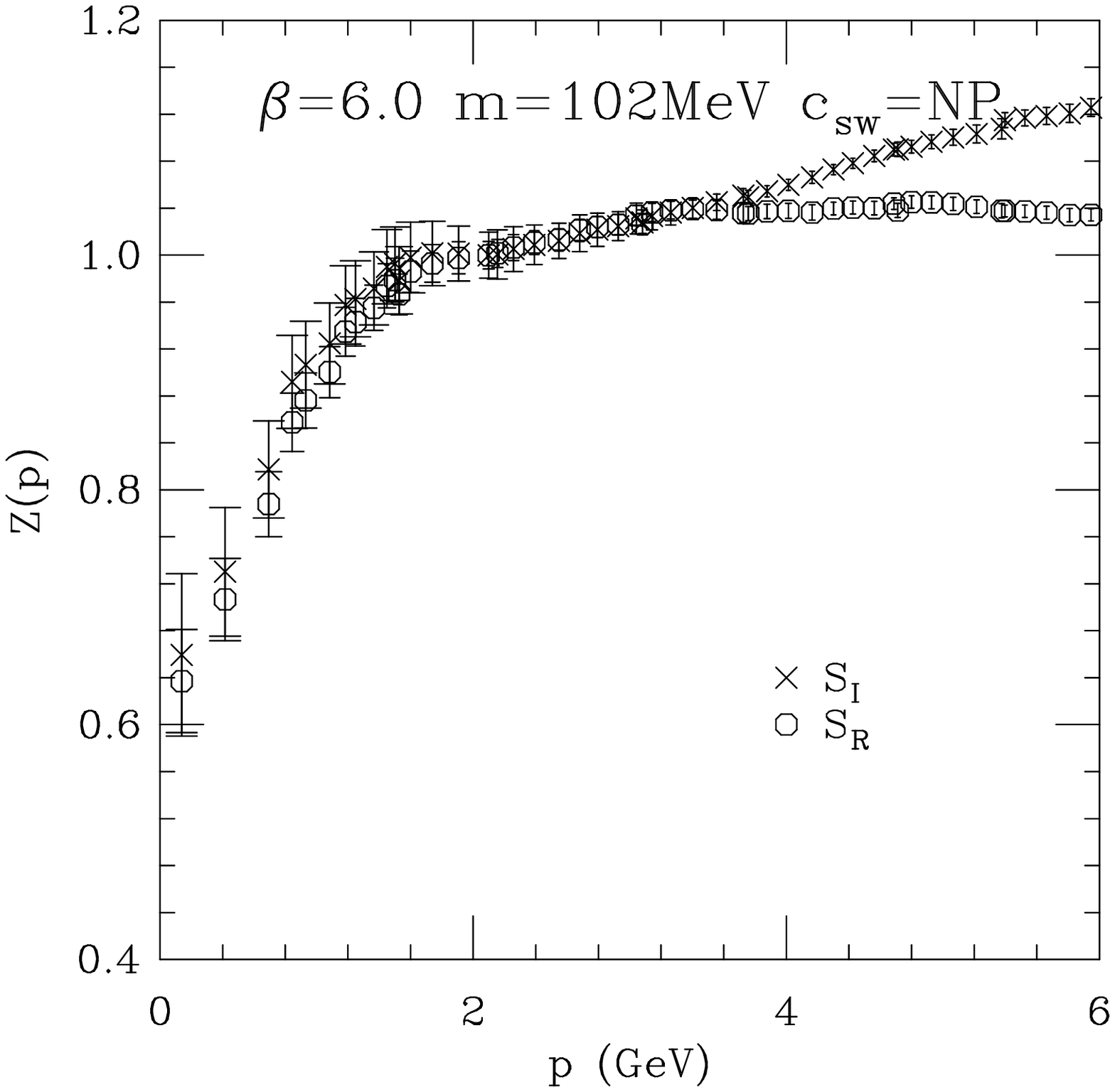}}\end{picture}}
\put(7,0){\begin{picture}(7,7)\put(-0.9,-0.4){\ebox{8cm}{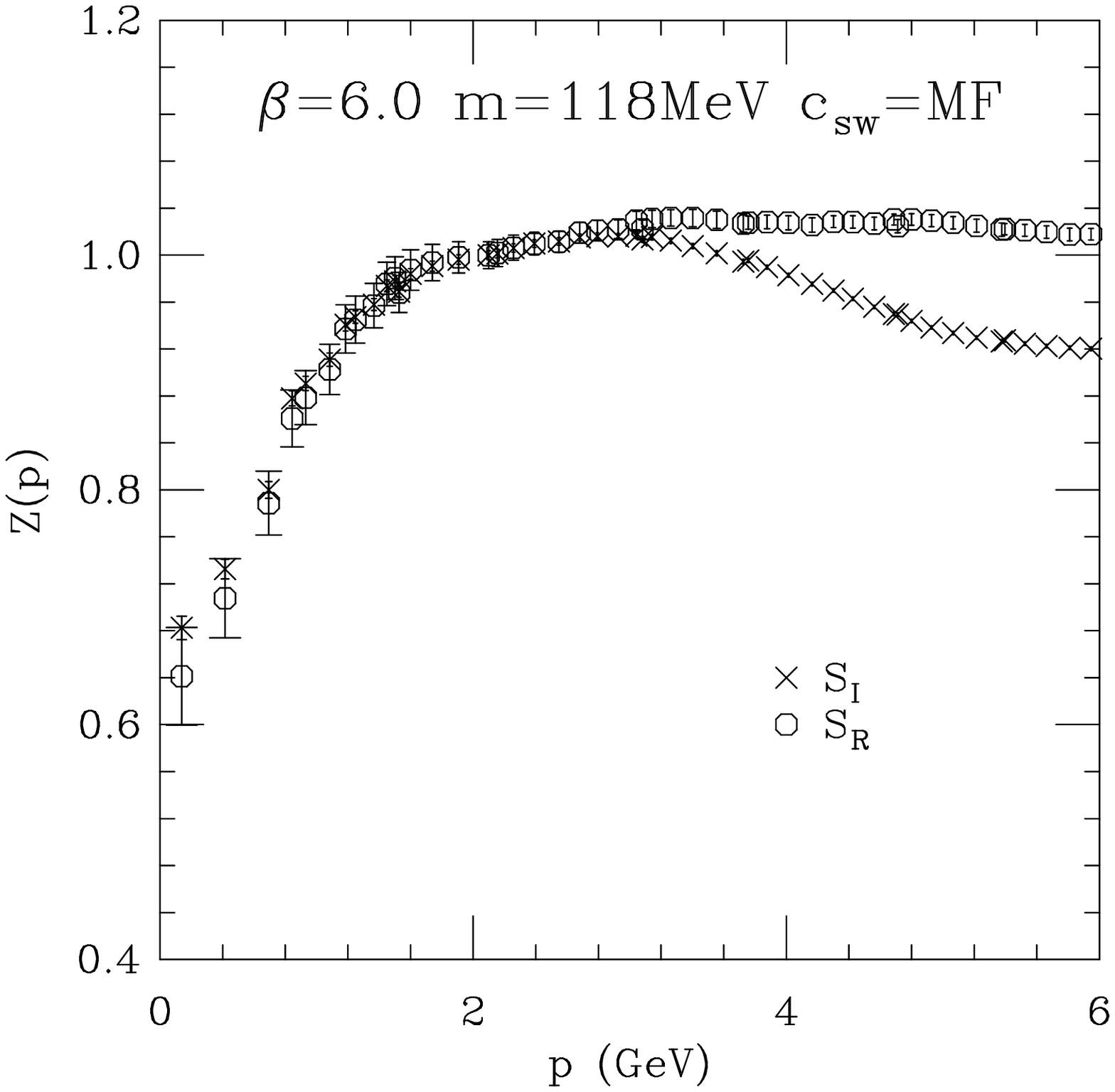}}\end{picture}}
\end{picture}
\end{center}
\caption{
The tree-level corrected $Z(p)$, for $\csw=\text{NP},
\kappa=0.137$ (left) and $\csw=\text{MF}, \kappa=0.13344$ (right).
The upper two figures show the data in lattice units, and without
rescaling.  The lower two figures show $Z(p)$ vs.\ momentum in
physical units, and after rescaling (``renormalizing'') so that
$Z(2.1 \text{GeV})=1$. 
The infrared agreement after rescaling is very good and we take the tree-level
corrected $Z(p)$ from $S_R$ with the nonperturbative $\csw$ as the best
estimate for this quantity (see text).
}
\label{fig:z_np_compare}
\end{figure}

In Fig.~\ref{fig:z_np_compare} we show the tree-level corrected $Z(p)$
function, for both nonperturbative and mean-field $\csw$.  The upper
figures show the `unrenormalized' $Z(p)$ --- the upper right figure is
taken directly from Fig.~5 in Ref.~\prev.  We see that there is still
a very significant discrepancy between $S_I$ and $S_R$, even with the
nonperturbative $\csw$.  Much of this discrepancy, however, amounts to
an overall renormalization, which may be included in the quark field
renormalization constant $Z_2$.  To eliminate this possible,
unphysical source of disagreement we rescale the data by imposing the
`renormalization condition' $Z(pa=1)=1$.  The result of this is shown
in the lower panel of Fig.~\ref{fig:z_np_compare}.
We then see that the infrared behavior of the
tree-level corrected $Z(p)$ functions agree much better than they did in
the previous work in Ref.~\prev.  
For the few most infrared points we see that there is an apparently
better agreement for $Z(p)$ between the two forms of the propagator
when the nonperturbative 
$\csw$ is used.  The ultraviolet agreement is less than satisfactory
even when the nonperturbative $\csw$ is used and we conclude that
the ultraviolet behavior of the tree-level form of $S_I$ is too severe to
be remedied by our tree-level correction scheme even with nonperturbative
and mean-field improved coefficients in the action and propagators
respectively.  Because of the more
reasonable tree-level behavior of $S_R$ and because in this case
$Z(p)$ is almost unchanged when using either the nonperturbative
or mean-field $\csw$, we take as our best estimate for $Z(p)$ the
tree-level corrected result from $S_R$ with nonperturbative $\csw$.

\begin{figure}
\begin{center}
\setlength{\unitlength}{1.1cm}
\setlength{\fboxsep}{0cm}
\begin{picture}(14,14)
\put(0,7){\begin{picture}(7,7)\put(-0.9,-0.4){\ebox{8cm}{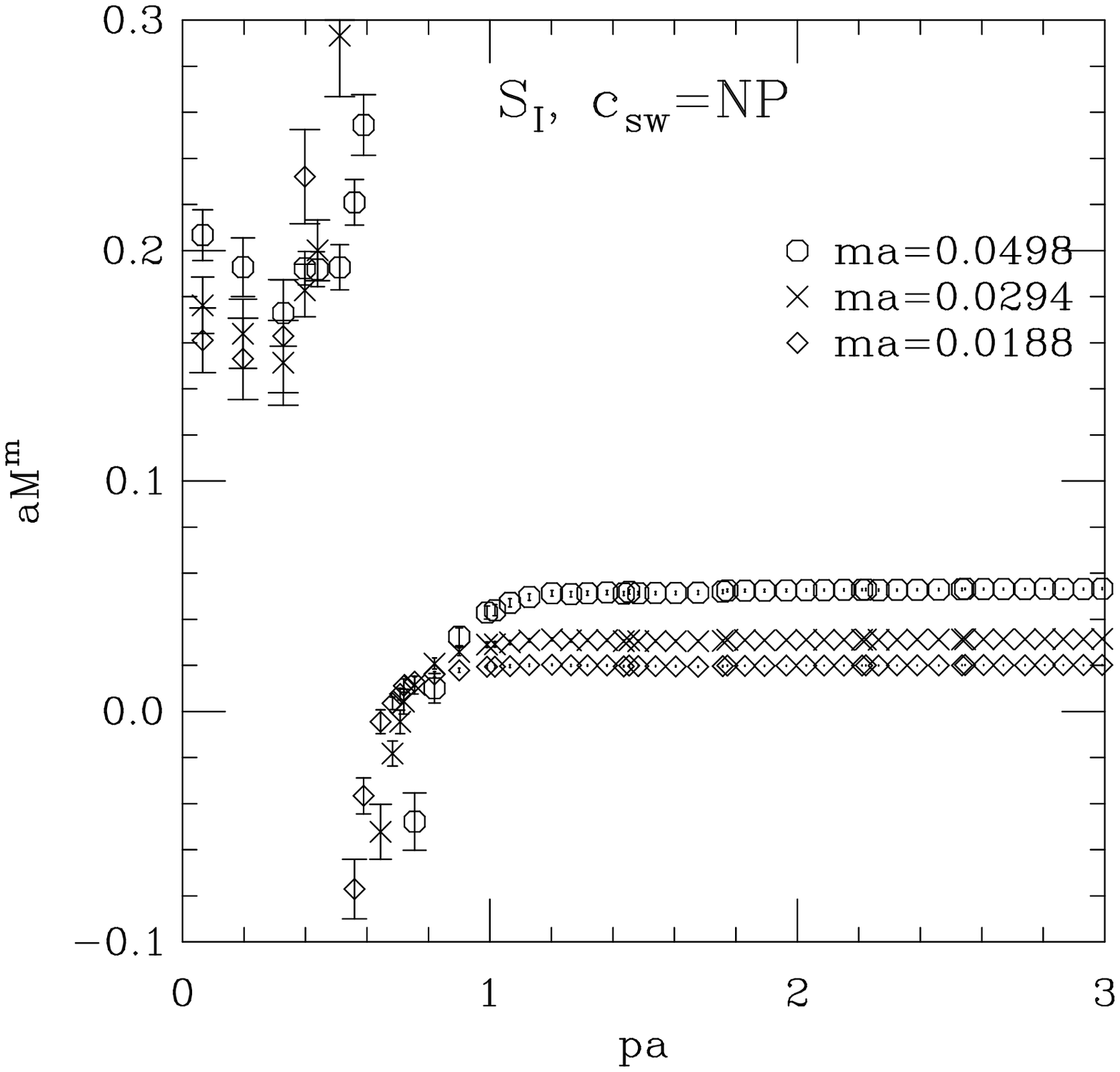}}\end{picture}}
\put(7,7){\begin{picture}(7,7)\put(-0.9,-0.4){\ebox{8cm}{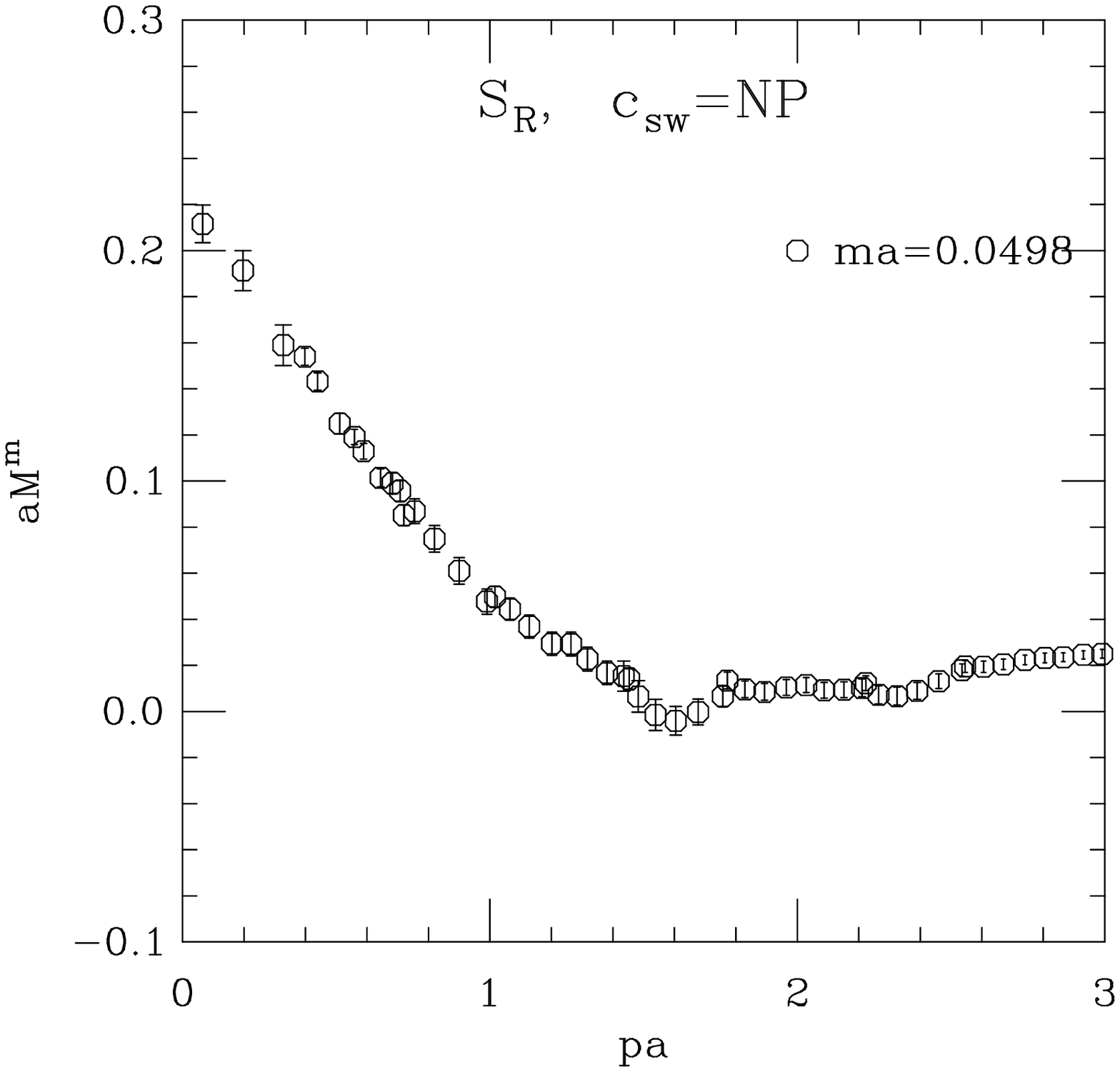}}\end{picture}}
\put(0,0){\begin{picture}(7,7)\put(-0.9,-0.4){\ebox{8cm}{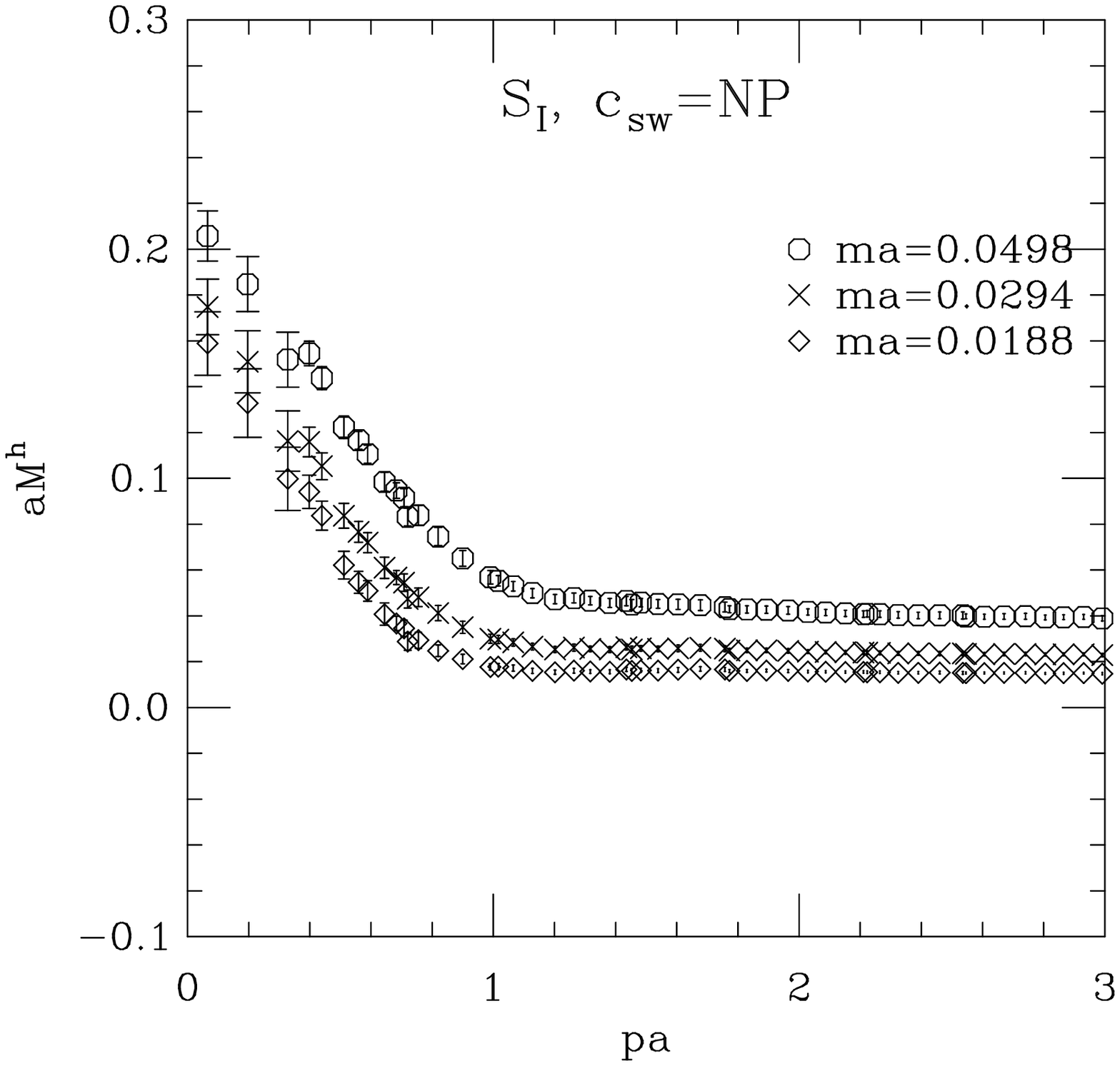}}\end{picture}}
\put(7,0){\begin{picture}(7,7)\put(-0.9,-0.4){\ebox{8cm}{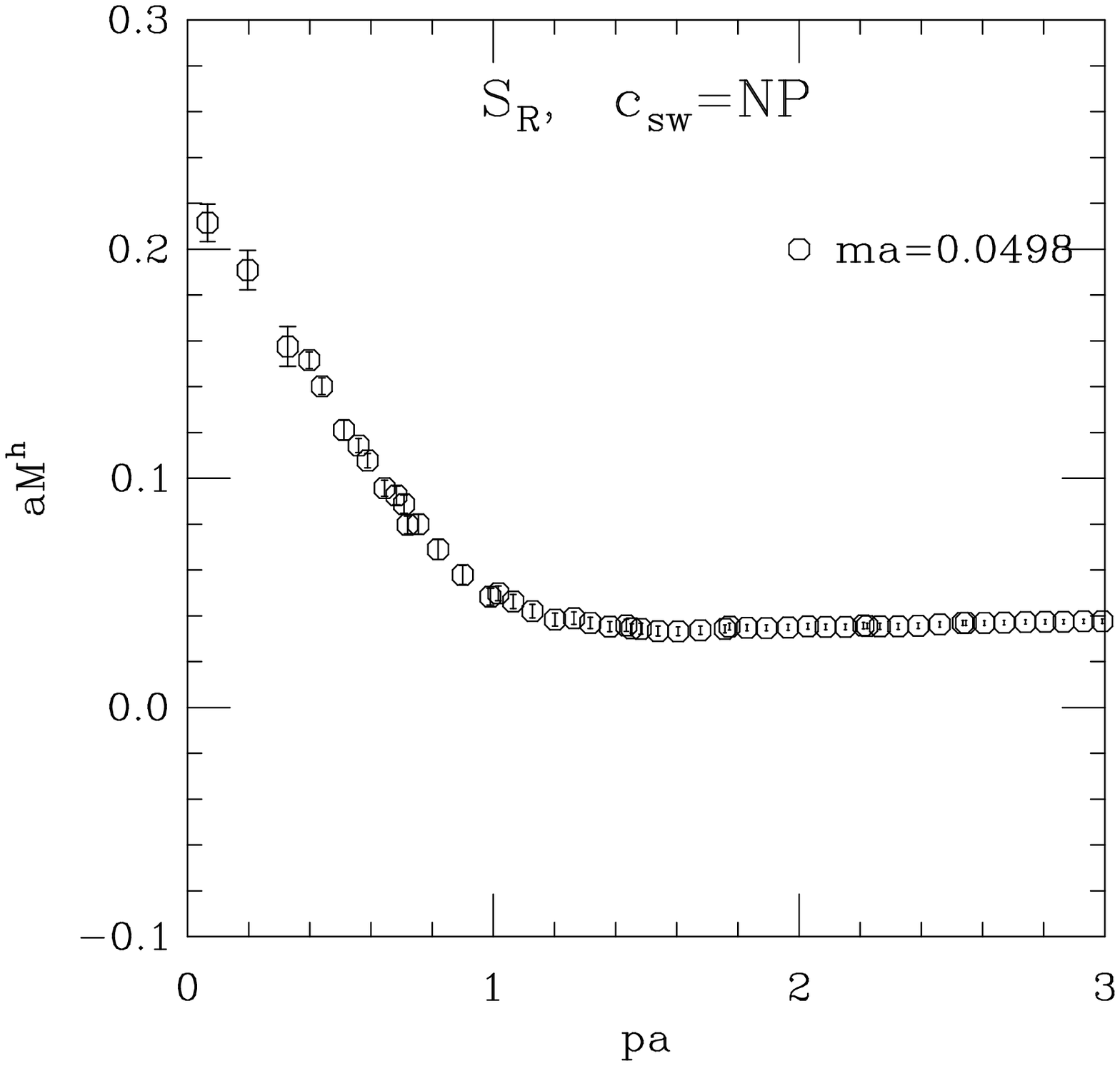}}\end{picture}}
\end{picture}
\end{center}
\caption{
The tree-level corrected $M(p)$, for $\csw=\text{NP}$, using
$S_I$ (left) and $S_R$ (right), and with the multiplicative (top) and
hybrid (bottom) correction schemes.  The multiplicative scheme clearly
fails for $S_I$, and also performs poorly for $S_R$; while the hybrid
scheme performs well for both and leads to good agreement between the
two propagators. 
}
\label{fig:m_np_schemecompare}
\end{figure}

\begin{figure}
\begin{center}
\setlength{\unitlength}{1.1cm}
\setlength{\fboxsep}{0cm}
\begin{picture}(14,7)
\put(0,0){\begin{picture}(7,7)\put(-0.9,-0.4){\ebox{8cm}{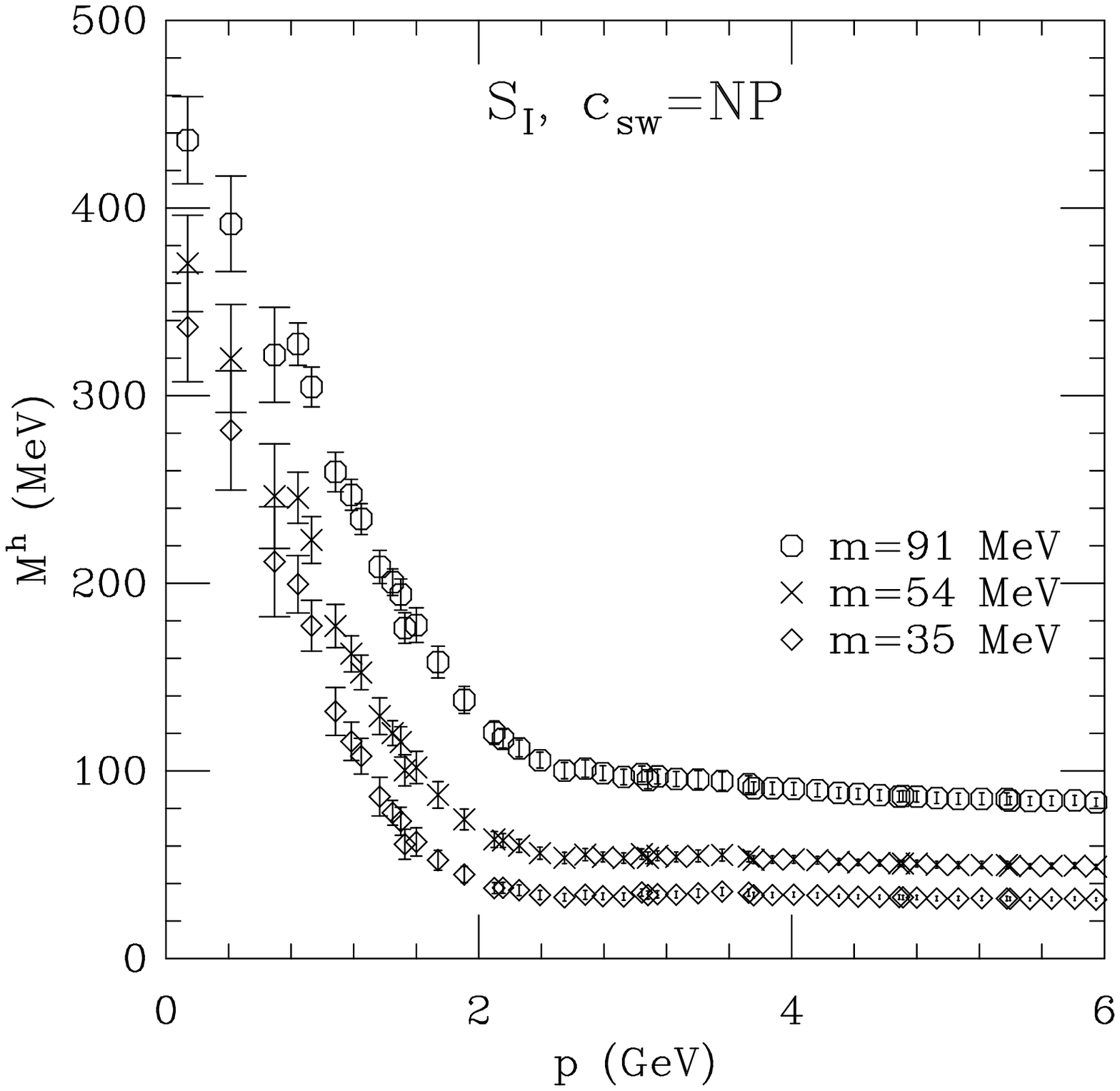}}\end{picture}}
\put(7,0){\begin{picture}(7,7)\put(-0.9,-0.4){\ebox{8cm}{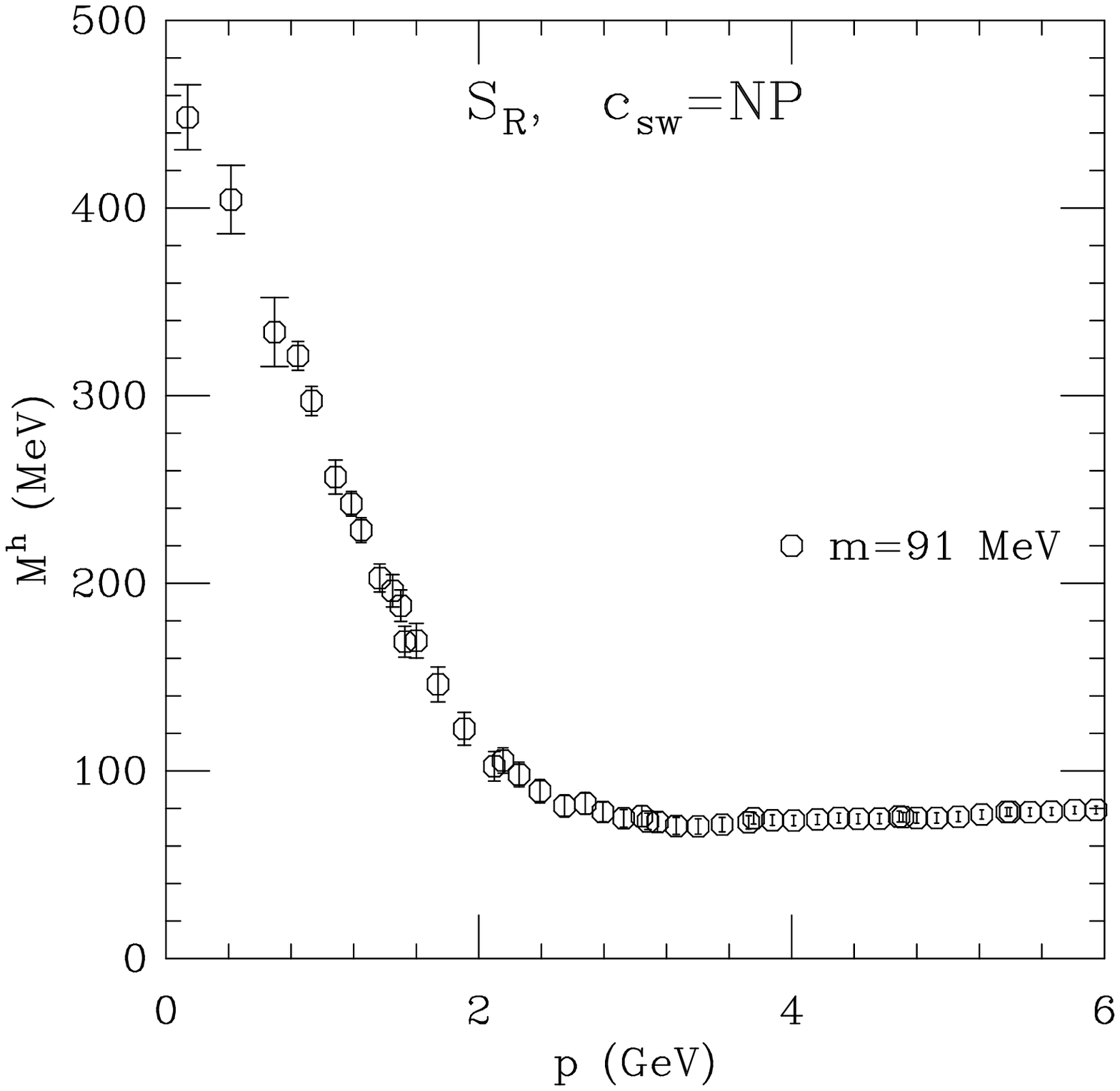}}\end{picture}}
\end{picture}
\end{center}
\caption{
The hybrid tree-level corrected $M^h(p)$, for $\csw=\text{NP}$, using
$S_I$ (left) and $S_R$ (right).  For $\mtil=91$ MeV we find good
agreement between the two data sets, both in the infrared and
ultraviolet.  The residual disagreement at intermediate momenta is a
pointer to lattice artifacts that we have not brought under full
control, even with nonperturbative improvement and hybrid tree-level
correction. 
}
\label{fig:m_np_phys}
\end{figure}

Figure~\ref{fig:m_np_schemecompare} shows the mass function for
$\csw=\text{NP}$, with the multiplicative and hybrid correction
schemes.  The multiplicative scheme exhibits the same problems as
those we encountered with $\csw=\text{MF}$.  In particular, $\zmz(p)$
from $S_I$ crosses zero for $pa\sim 0.5$ and renders the
multiplicative scheme meaningless for that case.  Even for $S_R$
we find that the uncorrected $M^L$ has small zero-crossings for
momentum values in the range $1.5<pa<2.3$, which render the multiplicative
scheme unsatisfactory.
Using the hybrid
scheme, however, we avoid these pathologies of the naive multiplicative
scheme as can be seen from the two lower figures in
Fig.~\ref{fig:m_np_schemecompare}.  In Fig.~\ref{fig:m_np_phys} we
have plotted the mass functions for the two propagators in physical
units.  We see good agreement between
the mass functions using hybrid tree-level correction and the nonperturbative
$\csw$ coefficient across the entire range of available momenta.
The mass function for $S_R$ dips slightly below that for $S_I$ at intermediate
momentum points even though they approach very similar asymptotic values.
This residual disagreement implies that we have not succeeded in removing
all of the lattice artifacts at intermediate momenta, although hopefully
we have gone some significant way toward achieving that end.

\subsection{Chiral extrapolations}

\begin{figure}
\begin{center}
\setlength{\unitlength}{1.1cm}
\setlength{\fboxsep}{0cm}
\begin{picture}(14,7)
\put(0,0){\begin{picture}(7,7)\put(-0.9,-0.4){\ebox{8cm}{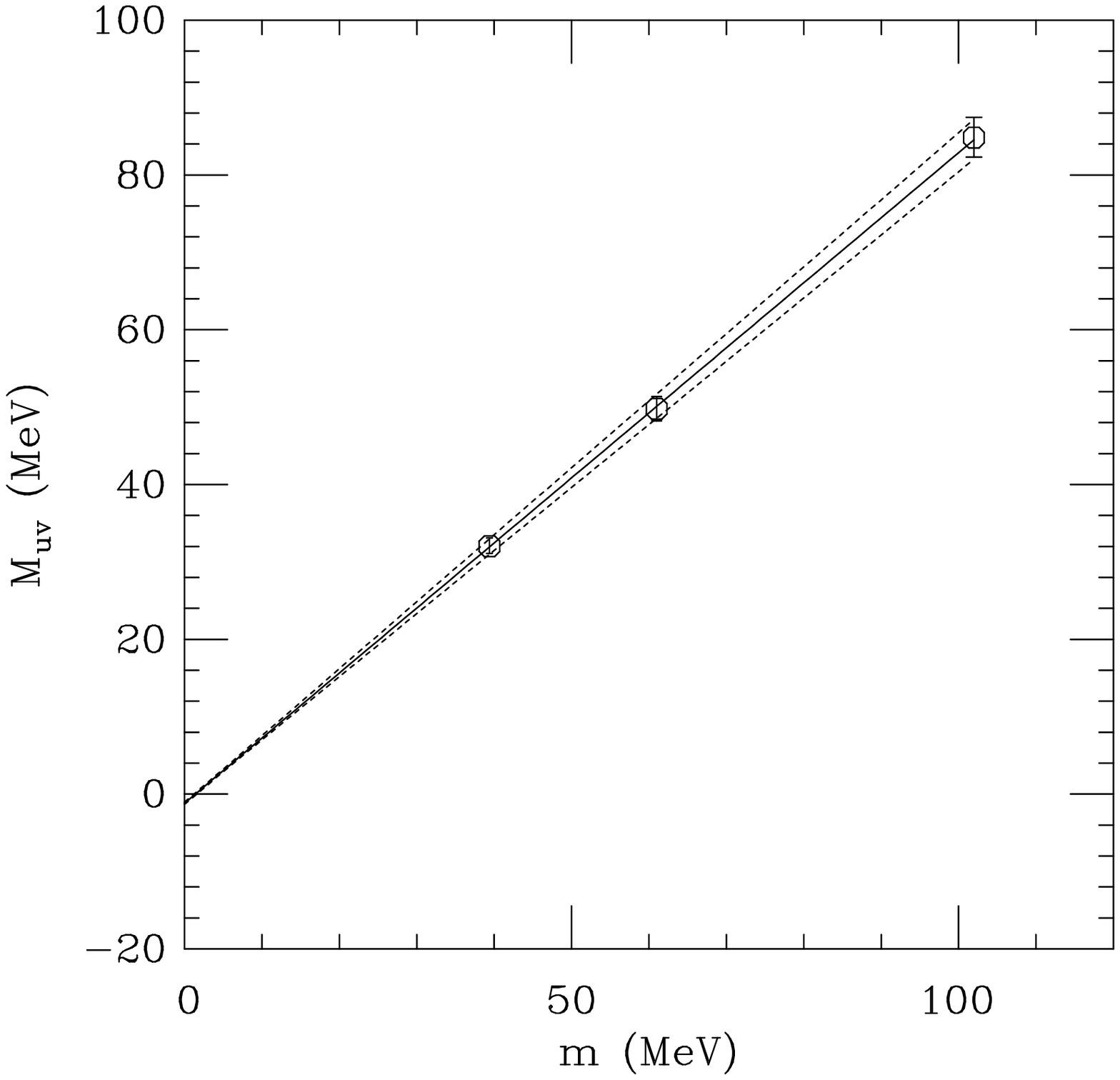}}\end{picture}}
\put(7,0){\begin{picture}(7,7)\put(-0.9,-0.4){\ebox{8cm}{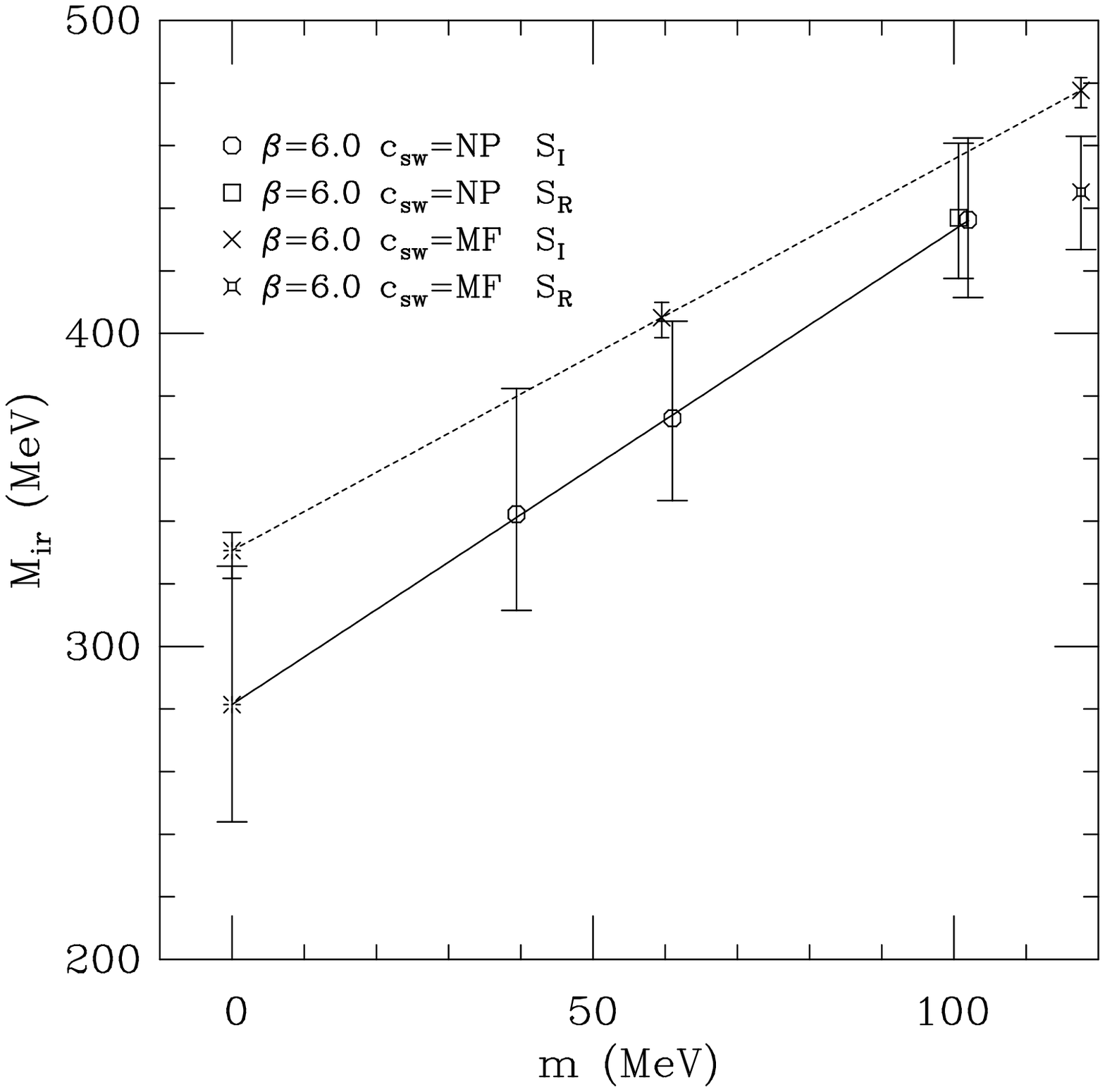}}\end{picture}}
\end{picture}
\end{center}
\caption{ Left: The ultraviolet quark mass $M_{\rm uv}$ as a function
of the bare quark mass $\mtil$, for $S_I$ with $\csw=\text{NP}$ and
using hybrid tree-level correction.  The values are obtained by
fitting $M^h(p)$ to a constant for $2<pa<3$.  Right: The infrared
quark mass $M_{\rm ir}\equiv M(p=0)$, obtained by extrapolating $M(p)$
to $pa=0$, as a function of $\mtil$.  The bursts indicate the chirally
extrapolated values of $M_{\rm ir}$ obtained by a simple straight line
fit for each action.  The solid line represents the fit for
$\csw=\text{NP}$, while the dotted line is the fit for
$\csw=\text{MF}$.  We see that the values of $M_{\rm ir}$ from $S_I$
and $S_R$ agree for $\csw=\text{NP}$, giving us further indication of
the superiority of nonperturbative to mean-field improvement, despite
the large statistical uncertainties in the NP data. }
\label{fig:M-chiral}
\end{figure}

We have available data for three quark masses for $S_I$ with the
nonperturbative $\csw$.   Having seen the very plausible behavior
of the mass function for $S_I$ after hybrid tree-level correction
and the good agreement with that for $S_R$, we are given the confidence
to attempt a simple extrapolation of the quark mass function to the
chiral limit.  The first chiral extrapolation we performed
was a linear extrapolation of the ultraviolet mass (obtained by fitting
$M^h(p)$ to a constant in the range $2<pa<3$) as a function of
$\mtil$.  This is shown in 
Fig.~\ref{fig:M-chiral}, where we see that the
ultraviolet mass vanishes in the chiral limit to a very good approximation
as it should.  
Indeed, the extrapolated value of $-1$ MeV is much
smaller than the systematic uncertainties arising from the different
tree-level correction schemes discussed above.
We also see that for the ultraviolet mass a linear
extrapolation does very well.
As we noted in Sec.~\ref{Sec:correct}, the ultraviolet mass
function is only constant up to logarithmic corrections,
but our ultraviolet behavior is not sufficiently under control
that it would be meaningful to attempt to extract those from
Fig.~\ref{fig:m_np_schemecompare}.

Also in Fig.~\ref{fig:M-chiral}, we show the result of a linear
extrapolation of the infrared quark mass $M(p\to0)$, together with our
previous results from Ref.~\prev.  Due to the small statistics, the
error bars for the nonperturbative $\csw$ data are quite large.  We
still observe that the
extrapolated mass value for the nonperturbative
$\csw$ is systematically lower than for the mean-field $\csw$, but the
results for $S_I$ and $S_R$ are now fully consistent, and also agree
with the value obtained from $S_R$ with the mean-field $\csw$.
Note that the value in MeV for $M(p\to0)$ from the mean-field $\csw$
differs from that reported in Ref.~\prev.  This is due to the
different values for the lattice spacing in the two papers.  The
uncertainty in the lattice spacing adds an additional uncertainty of
about 10\% to all numbers in physical units.  This uncertainty is an
intrinsic feature of the quenched approximation.

\begin{figure}
\begin{center}
\setlength{\unitlength}{1.1cm}
\setlength{\fboxsep}{0cm}
\begin{picture}(14,7)
\put(3.5,0){\begin{picture}(7,7)\put(-0.9,-0.4){\ebox{8cm}{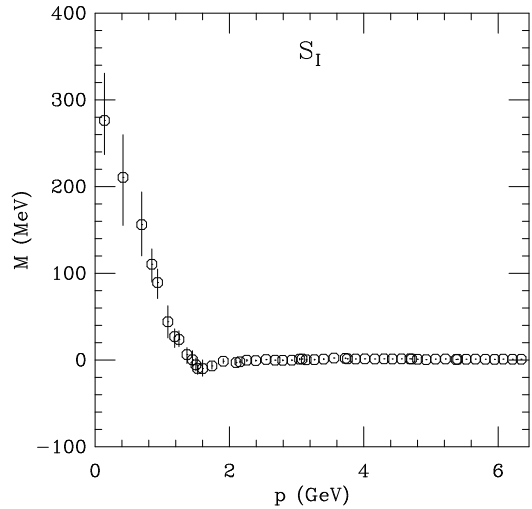}}\end{picture}}
\end{picture}
\end{center}
\caption{ The hybrid-corrected mass function from $S_I$ with
$\csw=\text{NP}$, with the bare mass $\mtil$ extrapolated to zero using a
quadratic fit.  The small dip at $p\sim1.6$ GeV is not statistically
significant and may be due to residual lattice artifacts.  The
non-zero values for $M(p)$ in the chiral limit are entirely due
to dynamical chiral symmetry breaking and provide a direct measure of
this effect. }
\label{fig:Mall-chiral}
\end{figure}

We also show in Fig.~\ref{fig:Mall-chiral} the result of a simple quadratic
chiral extrapolation in $\mtil$ for the entire mass function (for
$S_I$ with hybrid tree-level correction and nonperturbative $\csw$). 
The result has a plausible form and is presumably a good estimate
of $M(p)$ in the chiral limit.  The very small dip at $p\sim 1.4$~GeV
is within two standard deviations and, while not statistically
significant, it is a again a 
hint that we have not completely removed lattice artifacts at
intermediate momenta.  A linear chiral extrapolation, while adequate in the
ultraviolet and infrared, does not fit the data in the intermediate
momentum regime.  We see that the quadratic extrapolation to the chiral
limit is consistent with a vanishing current quark mass and a rapid
falling off of $M(p)$ such that it appears to essentially vanish by
approximately 1.5~GeV.  This suggests that the effects of dynamical
chiral symmetry breaking become negligible at a scale $p_\chi$ which
we estimate to be $p_\chi = 1.45\err{10}{13}$ GeV, where the errors
are purely statistical.

We have also studied the systematic uncertainties arising from the
specific choice of tree-level correction scheme.  In order to quantify
this, we have modified the hybrid scheme defined in
Sec.~\ref{Sec:correct} by taking
\begin{equation}
\dmp(pa) \to \dmp(pa) + \epsilon\dmm(pa) \qquad
\dmm(pa) \to (1-\epsilon)\dmm(pa)
\label{eq:hybrid-var}
\end{equation}
where $\epsilon$ is a free parameter.  Considering small variations in
$\epsilon$, $-0.1\lesssim\epsilon\lesssim0.1$, we find that the
correction scheme dependence gives rise to uncertainties in $p_\chi$
of about 100 MeV.  Using the mean-field $\csw$ at $\beta=6.0$
and 6.2, we get values for $p_\chi$ that are slightly higher, but
still consistent withint two standard deviations.  

We take as our best estimate for the chiral symmetry scale the value
from $S_I$ with the nonperturbative $\csw$ at $\beta=6.0$: 
$p_\chi=1.45\err{10}{13}\err{6}{8}(14)$ GeV, where the first set of
errors are statistical, the second are the systematic uncertainties
due to the tree-level correction scheme, and the third is the
uncertainty in the lattice spacing.  This value is roughly consistent
with the chiral symmetry breaking scale $\Lambda_{\rm{\chi}SB}$ arising
in low-energy effective theories and in instanton models (see e.g.\
Ref.~\cite{Schafer:1998wv}). 
An understanding of the relationship between this result
and the recent analysis of the pseudoscalar vertex \cite{Cudell:2001ny}
is an interesting topic for future investigation.

\section{Discussion}
\label{Sec:discuss}

We have made use of asymptotic freedom to factor out the dominant
(tree-level) lattice artifacts in the quark propagator at high
momenta.  We have discussed several different schemes for applying
this idea, referred to as tree-level correction, to the mass function,
which is the scalar part of the inverse quark propagator.  A purely
multiplicative scheme is seen to encounter problems because the value
of the tree-level mass function approaches or crosses zero, leading to
ill-defined behavior at intermediate momenta.  The purely additive
scheme defined in Ref.~\prev, although leading to a dramatic improvement on
the uncorrected data, did not give reliable results for the mass
function above $pa\sim1$.
We have defined a hybrid tree-level correction scheme which combines
the additive and multiplicative schemes in such a way that the mass
function becomes well-behaved at all momentum values.  Ambiguities in
the correction scheme should show up most clearly at intermediate
momenta.  We find that the uncertainties due to such ambiguities are
of comparable magnitude to the statistical uncertainties.

We have also studied the effect of using the nonperturbative value
for $\csw$.  This was seen to improve the data considerably, as
demonstrated most dramatically by the very good agreement between the
two definitions $S_I$ and $S_R$ of the improved quark propagator.  In
contrast, reducing the lattice spacing by going from $\beta=6.0$ to
$\beta=6.2$ with a mean-field $\csw$ only gave a slight
improvement. 

Finite volume effects may be estimated by studying the spread of
points with momenta in different directions in the infrared, as was
done in Refs.~\cite{Skullerud:2000un,Leinweber:1998uu}.  We do not
find any significant anisotropy at low momenta either in $Z(p)$ or in
$M(p)$, so we conclude that finite volume effects here are small.

We have not here considered the possible effect of Gribov copies.
This remains an interesting subject for future study, which is
currently being pursued.

In summary, the key results of this study are that one should use
the most appropriate (nonperturbative) determinations of improvement
coefficients wherever possible.  Where tree-level behavior is severe
with zero-crossings or near zero-crossings the hybrid
tree-level correction scheme can be used in place of the multiplicative
one.  In the infrared and intermediate momentum regimes we appear
to have reasonable control over lattice artifacts.  The chiral
behavior resulting from a simple chiral extrapolation appears
reasonable.  We believe that the best estimate of the
continuum $Z(p)$ function corresponds to the
nonperturbative $\csw$ result for the $S_R$ propagator as shown in
Fig.~\ref{fig:z_np_compare}. 
All of the hybrid corrected mass functions with nonperturbative
$\csw$ appear reasonable and agree with each other.  The chiral
extrapolation of these is shown in Fig.~\ref{fig:Mall-chiral}.  We
find that the effects of dynamical chiral symmetry breaking become
negligible above a momentum scale $p_\chi$.  Our best estimate for
this scale is $p_\chi=1.45\err{10}{13}\err{6}{8}(14)$ GeV.

We emphasize that the real test of these conclusions will be to
implement these methods on finer lattice spacings, with further
improved actions, and ideally with actions which respect chiral
symmetry on the lattice.  These subjects are being pursued.

\begin{acknowledgments} 

Financial
support from the Australian Research Council is gratefully
acknowledged.  The study was performed using UKQCD data obtained using
UKQCD Collaboration CPU time under PPARC Grant GR/K41663.  JIS
acknowledges support from the EU TMR network ``Finite temperature
phase transitions in particle physics'', EU contract ERBFMRX--CT97--0122.

\end{acknowledgments}


\end{document}